**Unconventional Spin-orbit Torques by Two-dimensional Multilayered MXenes for Future Nonvolatile Magnetic Memories**

*Prabhat Kumar,[1] Yoshio Miura,[1,2] Yoshinori Kotani,[3] Akiho Sumiyoshiya,[3] Tetsuya Nakamura,[3,4] Gaurav K. Shukla,[1] and Shinji Isogami [1]\**

[1] Research Center for Magnetic and Spintronic Materials,
National Institute for Materials Science (NIMS).
Sengen 1-2-1, Tsukuba, Ibaraki, 305-0047, JAPAN.

[2] Faculty of Electrical Engineering and Electronics,
Kyoto Institute of Technology.
Hashikami-cho, Matsugasaki, Sakyo-ku, Kyoto, 606-8585, JAPAN

[3] Photon Science Innovation Center (PhoSIC),
Aoba 468-1, Aramaki-Aza, Aoba, Sendai, 980-8572, JAPAN.

[4] International Center for Synchrotron Radiation Innovation Smart Center (SRIS),
Tohoku University.
Aoba 468-1, Aramaki-Aza, Aoba, Sendai, 980-8572, JAPAN.

*E-mail : isogami.shinji@nims.go.jp

**Abstract:**

MXenes have attracted considerable attention in recent years owing to their two-dimensional (2D) layered structures with various functionalities similar to those of graphene and transition metal dichalcogenides. To open a new application field for MXenes in the realm of electronic devices, such as ultrahigh-integrated magnetic memory, we have developed a spin-orbit torque (SOT) bilayer structure comprising bare MXene of $Cr_2N$: substrate//$Cr_2N$/[Co/Pt]$_3$/MgO using the magnetron sputtering technique. We demonstrated field-free current-induced magnetization switching (CIMS) in the bilayer structure, regardless of the charge current directions with respect to the mirror symmetry lines of $Cr_2N$ crystal. This is a specific characteristic for the 2D MXene-based SOT-devices, originating from an unconventional out-of-plane SOT. As the SOT efficiency increases with increasing the $Cr_2N$ thickness, the first-principles calculations predict an intrinsic orbital-Hall conductivity with the dominant out-of-plane component, comparing to the spin-Hall conductivity in the $Cr_2N$. X-ray magnetic circular dichroism reveals the out-of-plane uncompensated magnetic moment of Cr ($m_{Cr}^{UC.}$) in the $Cr_2N$ layer at the interface, induced by contact with the Co in the [Co/Pt]$_3$ ferromagnetic layer. Therefore, the intrinsic bulk orbital-Hall effect in MXene and the interfacial contribution such as spin-filtering-like effect owing to $m_{Cr}^{UC.}$ are considered as possible major mechanisms for the unconventional out-of-plane SOT in the device, rather than a crystal symmetry and/or an interlayer exchange coupling.

# 1. Introduction

The importance of semiconductor devices is rapidly increasing owing to the development of a modern society, in which humans are connected to all types of applications via the Internet. To realize efficient devices, two-dimensional (2D) materials, such as the single-layer graphene and the transition metal dichalcogenides (TMDC), have attracted significant attention owing to their various functionalities.[1-9] Furthermore, transition metal carbide $Ti_3C_2$ with atomic layered structures was first discovered in 2011,[10] and has opened a new class of 2D materials. It is termed as MXene and is also known as post-graphene and TMDC.[11] The chemical formula is $M_{n+1}X_nT_x$, where the site $M$ represents the transition metals, such as Ti and Cr, $X$ represents the 2$p$ light elements C or N, and $T$ represents surface terminations such as O and Cl on the outer $M$ layer. Specifically, $n$ varying from 1 to 4 corresponds to the number of the simplest $M$-$X$-$M$ bonding trilayer units of $M_2X$, while $x$ is a variable. The physical and chemical properties can be tailored by $n$, as well as by various combinations of $M$, $X$, and $T$.[12] These characteristics are related to the significant electronegativity of 2$p$ light element $X$, which allows for the strong orbital hybridization with the $M$ elements.[13] Thus, MXenes are widely considered to have immense potential as key materials for many device applications. In the past decade, MXenes have contributed to the fields such as biomedicine,[14] mechanical science,[15] optoelectronics,[16] and energy storage.[17] These pioneering works inspired our interest in finding more remarkable potential for MXenes. In this study, we aim to develop another application field of MXene by expanding it to the field of 2D spintronics,[18,19] which has been unfamiliar with MXene, and elucidated its superiority. Furthermore, we examined the MXene-specific spin-transport phenomena beyond charge transport.

In the spintronics research, manipulation of the magnetic moment via the spin degree of freedom has attracted considerable attention in terms of electronic devices with low power consumption because the spin current, a flow of spin angular momentum without electron charge, does not consume power in principle.[20] Specifically, to store the enormous amount of data associated with the widespread use of the Internet and mobile applications, nonvolatile spin-orbit torque magnetic random access memory (SOT-MRAM) has been extensively studied as one of the storage devices that takes advantage of the spin current.[21] Although three types of SOT-MRAMs with the magnetization directions of the recording layer in $x$, $y$ and $z$ have been proposed,[22] the so-called Type-Z with perpendicularly oriented magnetization (in $z$ direction) has been identified as a promising geometry for further high-integration and low-power MRAM packages. However, the Type-Z requires an in-plane bias magnetic field for writing using conventional in-plane current-induced SOT, originating from the spin current in conventional heavy metals, such as Ta, Pt, and W. This has been an issue for achieving the miniaturization of cell size and highly integrated SOT-MRAMs, leading to a recent challenge for field-free current-induced magnetization switching (CIMS) by SOT, and many approaches have been demonstrated to date:[23] the compositional and/or geometrical gradient in the multilayers;[24,25] the interfacial engineering by nonmagnetic layers;[26,27] the exchange coupling between antiferromagnet/ferromagnet bilayers;[28,29] the crystal symmetricity of noncollinear antiferromagnet and the low-symmetry TMDC;[30-37] and the complex circuit architecture with combined SOT and spin-transfer-torque.[38,39] These are based on the concepts of superimposing the out-of-plane SOT component on a conventional in-plane SOT, which is referred to as an unconventional SOT.

In addition to the aforementioned artificial structures and materials, we focus on the 2D bare MXene ($M_2X$) as a spin source layer to realize field-free CIMS. Although there are many candidates for the $M$ sites, such as Ti, V, W, Mo, and Cr, we employed $Cr_2N$ in terms of the phase stability and metallic conductivity, as in the case of conventional transition metal nitrides.[40,41] There are three aspects to be focused on for the MXene-based SOT-device. (i) A low-symmetry driven unconventional out-of-plane SOT, similar to the noncollinear antiferromagnet and TMDCs,[30-37] in which field-free CIMS can be induced by an in-plane charge current orthogonal to a mirror symmetry line of the crystal, but not in parallel. (ii) An orbital Hall effect (OHE) in the bulk part of MXene layer. The bilayer system with oxidized light elements exhibits an unconventional SOTs in spite of their weak spin-orbit interaction (SOI),[42-44] which is discussed based on the transfer of orbital angular momentum. (iii) An interfacial contribution to the out-of-plane SOT. Owing to the advantageous properties of 2D materials, the MXene exhibits an atomically flat interface with the ferromagnetic layer, enabling the discernment of interfacial effects such as electronic band structure and interlayer magnetic coupling. Recently, van der Waals 2D heterostructures enable field-free CIMS by the unconventional SOT,[45] which is discussed with the interfacial states such as efficient spin transparency and interfacial magneto-spin Hall effect.[46] Therefore, we have measured the CIMS with two different directions of in-plane charge current and various thickness of MXene, and we have quantified polarized spins at an interface via synchrotron radiation in this study.

In Fig. 1, we depict the possible interpretations of the field-free CIMS in MXene-based SOT-device based on the findings of this study. Firstly, the spin current in the $Cr_2N$ layer is expected to originate from the OHE with out-of-plane component, which dominates the entire SOT exerted on the ferromagnetic layer. Second, the nonmagnetic Cr layer, i.e. the top trilayer-unit of the $Cr_2N$ MXene adjacent to the ferromagnetic layer, can be polarized, resulting in an uncompensated magnetic moment of Cr ($m_{Cr}^{UC.}$) in the out-of-plane direction, as depicted by the thick arrows at the interface. This is primarily responsible for the superposition of the out-of-plane SOT component owing to a spin-filtering-like mechanism at the interface.[27,47] In another words, the spin converted from the orbital angular momentum in the same direction as the $m_{Cr}^{UC.}$, can be transferred, while the spin in the opposite direction cannot be transferred. This is a unique mechanism for the field-free CIMS in the MXene-based SOT-devices, which cannot be explained by the existing scenario of 2D crystal symmetry.

## 2. Results and discussion

The unit cell of Cr$_2$N MXene shows a hexagonal structure with a space group of $P\bar{3}1m$, as shown in Fig. 2(a1), and the lattice constants are $a = b = 0.48$ nm and $c = 0.45$ nm. The collinear antiferromagnetic structure has been reported for a wide temperature range from 100 K to 500 K,[48-52] which was also predicted by our first-principles calculations, as indicated by the arrows on each Cr atom. The cross-sectional and plane views are shown together with their coordinates in Figs. 2(a2) and 2(a3), where the Cr atoms in the top layer are surrounded by the black circles to distinguish them from the bottom Cr layer. The top and bottom Cr layers exhibited a close-packed structure, and N atoms were present at the octahedral sites between the two Cr layers. Note that the present Cr$_2$N belongs to the family of bare MXene without $T$ sites, but the Cr-N-Cr trilayer unit is bonded via stacking of N layers. Figure 2(b1) shows the out-of-plane X-ray diffraction (XRD) profiles of pure Cr, CrN, and Cr$_2$N films with a thickness of ~20 nm on the $c$-plane oriented Al$_2$O$_3$ substrate. Each phase can be formed using different ratios of N$_2$ gas flow, defined as $Q$ = N$_2$/(Ar + N$_2$), in reactive sputtering deposition. At $Q = 0\%$, a pure Cr film was grown with (110) texture. For $Q = 5\%$, the fringe oscillation was observed near the XRD peak at $2\theta/\omega \approx 40°$ and 88°, suggesting the Cr$_2$N-MXene phase formation with hexagonal crystal structure and atomically flat interfaces. For $Q = 10\%$, the CrN, which is another phase of the Cr-N intermetallic compound with face-centered-cubic structure, was grown in (111) texture. Figure 2(b2) shows the substrate temperature ($T_{sub}$) dependence of the out-of-plane XRD profiles of Cr$_2$N films with the same thicknesses. Although the texture with (0001) orientation can be grown even at room temperature (RT), high atomic order was obtained at $T_{sub}$ ranging from 350 °C to 650 °C. Figure 2(c1) shows the high-resolution cross-sectional transmission microscopy (TEM) images of Cr$_2$N deposited at 650 °C. Figure 2(c) shows the atomic force microscope image for the 20-nm-thick Cr$_2$N film to show the long range flatness. The root mean square roughness was evaluated to be only 0.12 nm, suggesting the atomically flat surface as indicated by the fringe oscillation in Fig. 2(b1). Figures 2(d) and 2(e) show the in-plane magnetic properties (in $x$-direction) and the anomalous Hall measurements with the magnetic field along the out-of-plane direction (in $z$-direction) of Cr$_2$N layer. The magnetization was negligible without hysteresis, which can be attributed to the antiferromagnetism of Cr$_2$N, as predicted by first-principles calculations.[48-52]

To investigate the CIMS characteristics owing to Cr$_2$N MXene, we prepared SOT-devices, as depicted in Fig. 3(a): substrate//Cr$_2$N(5 nm)/[Co(0.35 nm)/Pt(0.3 nm)]$_3$/MgO(2 nm). The Co/Pt multilayer with three periods, which is described as [Co/Pt]$_3$, provides sufficient perpendicular magnetic anisotropy, resulting in an out-of-plane magnetization ($M_{Co/Pt}$) at the remanent state (see Fig. S1 in the Supporting Information). Figures 3(b1) and 3(b2) exhibit (0001) plane of the Cr$_2$N supercell, which is the same as that shown in Fig. 2(a3), and mirror symmetry axes are depicted by the blue dashed lines ($m$). In order to explore the specific characteristics of Cr$_2$N/ferromagnet system that cannot be explained by such mirror symmetry, we intentionally applied a charge current pulse parallel and orthogonal to the mirror axis, in which the out-of-plane SOT vanished for only the configuration of parallel [Fig. 3(b1)] in the case of WTe$_2$/ferromagnet systems.[33,34] The anomalous Hall resistance ($R_{xy}$) with the magnetic field sweep along the out-of-plane ($H_z$) and in-plane ($H_x$) direction is shown in Figs. 3(c1) and 3(c2), respectively, where $M_{Co/Pt}$ is indicated by arrows. The $R_{xy}$ amplitude for $\mu_0H_z = 0$ was consistent with that for $\mu_0H_x = 0$, suggesting sufficient perpendicular magnetic anisotropy in [Co/Pt]$_3$ layer such that $M_{Co/Pt}$ points in $z$-direction at the magnetic remanent state. Figure 3(c3) shows the CIMS behaviors under the bias field of $H_x$, where the load current pulse duration is 10 ms and the direction is parallel to $m$ ($m \parallel I$) as described in Fig. 3(b1). The change in $R_{xy}$, corresponding to the magnetization switching, was remarkably sharp as observed in the conventional heavy-metal based SOT-device with Type-Y geometry,[53] and the polarity of CIMS loop depending on the $H_x$ was clockwise (CW) for $\mu_0H_x = +29$ mT and counter-clockwise (CCW) for $\mu_0H_x = -18$ mT. It should be noted that the partial CIMS occurred at $\mu_0H_x = 0$ (field free) as shown in Fig. 3(c4), and the effective critical current density flowing in the Cr$_2$N layer ($J_{Cr2N}$) was ~30 MA/cm$^2$. Specifically, the value was obtained by eliminating the current shunting to the [Co/Pt]$_3$ layer based on the measured resistivity of Cr$_2$N and [Co/Pt]$_3$ layers (see Fig. S2 in the Supporting Information). Note that the $J_{Cr2N}$ obtained is comparable to that of Type-Y devices,[53] which include heavy-metals with strong SOI and high SOT efficiency. The polarity of the field-free CIMS was CW, which was confirmed for the other 10~20 devices. The $J_{Cr2N}$-$H_x$ diagram shown in Fig. 3(c5) revealed that the $J_{Cr2N}$ decreased with increasing $H_x$, and the field-free CIMS was achieved even at lower $T_{sub} = 350$ °C (see Fig. S3 in the Supporting Information). In the production line of the CMOS transistor, in which the SOT-MRAMs are to be embedded, post-annealing has been performed at ~400 °C. Note that the Cr$_2$N based SOT-devices have enough heat endurance to exhibit a stable CIMS owing to the robustness of the Cr$_2$N crystal structure with respect to growth temperature [Fig. 2(b2)]. The ratio of field-free CIMS to the full CIMS was approximately 20%, regardless of $T_{sub}$, as shown in Fig. 3(c6). These intermediate states were also observed in the other devices as well we measured (see Fig. S4 in the Supporting Information). Nevertheless, the ratio can be enhanced by the 1-nm-thick Cr insertion (blue symbol), while it is slightly suppressed by the 1-nm-thick Pt insertion (green symbol), suggesting a relationship between the interfacial phenomena and out-of-plane SOT component. Given that [Co/Pt]$_3$ ferromagnetic layer of the Hall-cross device has a large area with 10 μm × 7 μm, multiple domain structures would form when CIMS occurs, resulting in an intermediate magnetic state.[54] On the other hand, Fig. 3(d1) shows the field-free CIMS loops for the same SOT-devices with [Co/Pt]$_3$ ferromagnetic pillar with ~7 μm in diameter. Note that the in-plane charge current flowed along the mirror symmetry line, in which out-of-plane SOT is not allowed in principle. The switching ratio was much larger than that with Hall-cross geometry [Fig. 3(c4)], leading to ~94 % of full switching amplitude (see Fig. S5 in the Supporting Information). Figure 3(d2) shows the CIMS with the in-plane charge current orthogonal to the mirror line, in which the out-of-plane SOT is allowed in principle. The field-free CIMS was also observed with similar switching ratio, suggesting an unconventional out-of-plane SOT that cannot be explained by the existing symmetry driven out-of-plane SOT mechanism.

To assess the possible mechanism of the unconventional SOT, the effective efficiency ($\xi_{eff}$) was quantified for the Hall-cross devices with various Cr$_2$N thicknesses: substrate//Cr$_2$N($t_{Cr2N}$)/[Co(0.35 nm)/Pt(0.3 nm)]$_3$/MgO(2 nm). Figures 4(a1) and 4(a2) show the out-of-plane AHE and CIMS loops, respectively, where the charge current direction was parallel to the mirror symmetry axis ($m \parallel I$). The amplitude of $R_{xy}$ decreased with increasing current-shunting into the Cr$_2$N layer; thus, we multiplied the factors to expand the loops for visibility. The amplitudes of AHE and CIMS were comparable for each $t_{Cr2N}$, and the critical current density of $J_{Cr2N}$ decreased with increasing $t_{Cr2N}$. The $\xi_{eff}$ was estimated using the following equation,[55,56]

$$\xi_{eff} = \left(\frac{2e}{\hbar}\right)\left(\frac{M_s t_{Co/Pt} H_p}{J_{Cr_2N}}\right) \quad , \tag{1}$$

where $M_s$ and $H_p$ denote the saturation magnetization of Co/Pt ferromagnetic layer and domain wall depinning field that is defined as $H_c = H_p/\cos\theta$, respectively.[55,56] Given that the CIMS in the present Hall-cross structure occurs through the domain nucleation and propagation process, the $H_p$ value is necessary to estimate $\xi_{eff}$ as shown in Eq. (1). $H_p$ was determined by measuring the $H_c$ as a function of the polar angle ($\theta$) with respect to the film surface (see Fig. S6 in the Supporting Information). Note that the current-shunting into [Co/Pt]$_3$ layer was excluded from $J_{Cr2N}$ by multiplying the ratio of the sheet resistance of the Cr$_2$N layer to the entire sheet resistance. It was revealed that $\xi_{eff}$ increased monotonically with increasing $t_{Cr2N}$ as shown by the red symbols in Fig. 4(c). To verify this thickness dependence, we examined $\xi_{eff}$ by another method, that is a second-harmonic SOT measurement. Figures 4(b1) and 4(b2) show the first- and second-harmonic Hall resistances, respectively, of the sample with $t_{Cr2N}$ = 3.6 nm. To distinguish between the damping- and field-like SOTs and various magneto-thermoelectric components,[57] such as the ordinary Nernst effect and anomalous Nernst effect, we conducted the data fitting using the following equation,[57]

$$R_{xy}^{2\omega} = \frac{R_{AHE}}{2}\frac{H_{DL}}{|H_x|-H_k^{eff}} + R_{PHE}\frac{H_{FL+Oe}}{|H_x|} + R_{TH} \,, \qquad (2)$$

at the high-field regime, as shown in Fig. 4(b2). $R_{AHE}$, $R_{PHE}$, and $R_{TH}$ denote the amplitudes of the anomalous Hall resistance, planar-Hall resistance obtained by the magnetization rotation on the *xy*-plane, and the resistance originating from the magneto-thermoelectric effect, respectively. $H_k^{eff}$ denotes the effective anisotropy field estimated using Eq. (S2) in the Supporting Information. $H_{DL}$ and $H_{FL+Oe}$ are the damping-like SOT effective field and the superposition of the field-like effective field and Oersted field, respectively. Using $R_{AHE(PHE)} \approx$ 7.5 Ω (0.1 Ω) and $\mu_0 H_k^{eff} \approx$ 0.6 T, we obtained $\mu_0 H_{DL} \approx$ 4.2 mT and $\mu_0 H_{FL+Oe} \approx$ 0.30 mT, resulting in $\xi_{eff} = \mu_0 H_{DL}/J_{Cr2N} \approx$ 0.66 mT/MA cm$^{-2}$ for $t_{Cr2N}$ = 3.6 nm. Note that $R_{TH}$ corresponds to the offset with respect to $R_{xy}^{2\omega} \approx 0$ at high field (~2 T) [Fig. 4(b2)], suggesting that the magneto-thermoelectric effect can be ruled out from the major origin for CIMS in the present Cr$_2$N/[Co/Pt]$_3$ system. These results demonstrate that $\xi_{eff}$ dependence on $t_{Cr2N}$ via second-harmonic measurements was in agreement with that obtained via CIMS, as shown in Fig. 4(c). Furthermore, we also evaluated the out-of-plane and in-plane SOT via low field second-harmonic Hall measurements using the sample with different ferromagnetic layer: substrate//Cr$_2$N(8.8 nm)/CoFeB(1 nm)/MgO(2 nm) (see Fig. S7 in the Supporting Information). The out-of-plane and in-plane $\mu_0 H_{DL}/J_{Cr2N}$ was ~0.18 mT/MA cm$^{-2}$ and 0.039 mT/MA cm$^{-2}$, respectively, indicating the presence of out-of-plane SOT based on the Cr$_2$N.

To explore the relationship between the SOT and the interface phenomena, we measured the element-selective magnetic properties at the Cr$_2$N/ferromagnetic interfaces by means of X-ray magnetic circular dichroism (XMCD) for three samples [Figs. 5(a1)-5(a2)]: (a1) the pure Co for comparison, (a2) Cr$_2$N/Co bilayer, and (a3) Cr$_2$N/Pt bilayer. The XMCD signals are shown in Figs. 5(b)-5(d), and Table 1 summarizes spin ($m_{spin}$) and orbital ($m_{orb}$) magnetic moments of Co, Cr, and N, estimated using the sum rule (see Fig. S8 in the Supporting Information).[58,59] The resultant values of pure Co ($m_{spin} \approx 1.83$ $\mu_B$; $m_{orb} \approx 0.17$ $\mu_B$) were consistent with the calculation results of $m_{spin} \approx 1.63$ $\mu_B$ and $m_{orb} \approx 0.1$ $\mu_B$,[60] while those of Cr$_2$N/Co ($m_{spin} \approx 0.99$ $\mu_B$; $m_{orb} \approx 0.073$ $\mu_B$) were smaller when compared to the pure Co. On the other hand, Cr in the Cr$_2$N/Co was clearly polarized to be $m_{spin} \approx -0.063$ $\mu_B$ as shown in Fig. 5(c), which are close to the values for Cr$_2$O$_3$/Co systems reported.[61] Conversely, $m_{spin(orb)}$ of Cr in Cr$_2$N/Pt was negligible. These results show the presence of $\boldsymbol{m_{Cr}^{U.C.}}$ originating from the imbalance in the antiferromagnetic structure of Cr$_2$N due to the adjacent Co layer, as shown in Fig. 1. Regarding N polarization, however, $m_{spin(orb)}$ was negligible, as shown in Table 1 and Fig. 5(d), while N in the Fe-N system shows a finite $m_{spin(orb)}$.[62] These results confirm the atomic-layered structure of Cr$_2$N, which is terminated by the Cr atomic-layer at the Cr$_2$N/Co interface. To provide further insights into the $\boldsymbol{m_{Cr}^{U.C.}}$, we measured the element-selective magnetic hysteresis loops along the $H_z$ direction for Co $L_3$-edge and Cr $L_3$-edge as shown in Fig. 5(e). Sharp magnetization switching of Co and Cr was evident at the same $H_z$, suggesting that the magnetic easy-axis was aligned in the out-of-plane direction. The switching directions were opposite to each other with respect to the magnetic field, indicating antiferromagnetic coupling between Co and Cr, i.e., $\boldsymbol{m_{Cr}^{U.C.}}$ points down (up) when the magnetization of Co points up (down).

**Table 1.** Spin magnetic moment ($m_{spin}$) and orbital magnetic moment ($m_{orb}$) for Co, Cr, and N, which are estimated using the Sum rule (see Fig. S8 in the Supporting Information).[58]

| Sample [$\mu_B$/atom] | Co $m_{spin}$ / $m_{orb}$ | Cr $m_{spin}$ / $m_{orb}$ | N $m_{spin}$ / $m_{orb}$ | Co (calc) a) $m_{spin}$ / $m_{orb}$ |
|---|---|---|---|---|
| Pure Co | 1.83 / 0.17 | N.A. | N.A. | 1.63 / ~0.1 |
| Cr$_2$N / Co | 0.99 / 0.73 | −0.063 / ~0 | ~0 / ~0 | N.A. |
| Cr$_2$N / Pt | N.A. | ~0 / ~0 | ~0 / ~0 | N.A. |

a) The values are referred to Ref. [59].

Hereafter, we discuss possible unconventional SOT mechanisms occurring in the Cr$_2$N/[Co/Pt]$_3$ system based on theoretical calculations and control samples. The SOT in a ferromagnetic layer generally originates from the spin current generated not only at the bulk part of the spin-source layer but also at the interface; therefore, both cases are considered individually. First, the spin diffusion length of pure Cr ($\lambda_s^{Cr}$) is reported as ~2.1 nm at RT and ~4.5 nm at 4.2 K.[63,64] Assuming $\lambda_s^{Cr2N} \leq \lambda_s^{Cr}$, due to high atomic density in Cr2N comparing to the pure Cr,[63] the $\xi_{eff}$ is expected to decrease for the thickness $t_{Cr2N} > \lambda_s^{Cr2N}$, if the spin current predominantly flows though the Cr2N layer. Given this hypothesis is not applicable, as indicated in Fig. 4(c), it is inferred that the spin current originating from the conventional spin-Hall effect at the bulk part of the Cr2N layer would be a minor cause. Instead, we must consider the long-range transport property that gives rise to an enhanced $\xi_{eff}$ for thicker $t_{Cr2N}$. Specifically, the OHE,[65] which is reported to emerge in the light elements with weak SOI, has longer orbital diffusion length ($\lambda_o^{Cr} \approx 6.1$ nm) comparing to the $\lambda_s^{Cr}$.[66] Furthermore, enhanced $\xi_{eff}$ by increasing the Cr thickness in the Co/Cr system has been reported by another group, which has been explained by the OHE.[66,67] In addition to such experimental results, we calculated the spin-Hall conductivity ($\sigma_{xz}^{spin(k)}$) and orbital-Hall conductivity ($\sigma_{xz}^{orb(k)}$) in the Cr2N [Figs. 6(a1) and 6(a2)] to enable the quantitative comparison between the SHE and OHE contributions in the Cr2N, where *x*, *z*, and *k* represent the directions of charge current, spin current, and spin/orbital polarization, respectively. Note that we focused only on the possible spin/orbital current flowing in *z*-direction, which contributes to CIMS of ferromagnetic layers. In addition, other components are summarized in Fig. S9 in the Supporting Information. Overall, $\sigma_{xz}^{spin(k)}$ was one or two orders of magnitude smaller than $\sigma_{xz}^{orb(k)}$, regardless of *k* direction. Therefore, the hypothesis of the dominant OHE that drown out from the experiments in Fig. 4(c) can be supported by the theoretical prediction, which is similar to the case of pure Cr: $\sigma^{spin} \approx -100$ ($\hbar/e$)(S/cm) and $\sigma^{orb} \approx 8000$ ($\hbar/e$)(S/cm).[68] Focusing on the *k*-dependence of OHE, note that we find $\sigma_{xz}^{orb(z)} > \sigma_{xz}^{orb(x)} > \sigma_{xz}^{orb(y)}$ at the Fermi level as shown in Fig. 6(a2). This implies that the orbital current with *z*-polarization (*k* = *z*) emerges in the bulk part of Cr2N in principle, which is converted into the spin current, resulting in the out-of-plane SOT for the field-free CIMS. To provide insight into the spin and orbital Hall conductivities, the spin and orbital Berry curvatures are analyzed as shown in Fig. S10 in the Supporting Information, together with the orbital projected band dispersion of Cr in Cr2N for the high-symmetry line. It was revealed that the *Γ-K* symmetry line of the Cr2N MXene dominates the contribution to the spin and orbital Hall conductivity. These characteristics are different from the conventional heavy metals, for example, high Berry curvature near the *X* and *L* points, and near the *P* point and along the path from *H* point contribute to the significant spin-Hall conductivity of the fcc Pt, and *α*-Ta, respectively.[69] The spin Hall conductivity at the Fermi Energy was ~2200 ($\hbar/e$)(S/cm) for the fcc Pt and ~ -142 $(\hbar/e)$(S/cm) for the $\alpha$-Ta.[69] $\sigma_{xz}^{\mathbf{orb}(k)}$ for the Cr$_2$N MXene was comparable to the value for the fcc Pt, while $\sigma_{xz}^{\mathbf{spin}(k)}$ was much smaller than those for the fcc Pt and the $\alpha$-Ta.

Next, it is essential to consider the interfacial contribution to the out-of-plane SOT, which can dominantly contribute to the field-free CIMS. Based on the XMCD results, we identified the $\boldsymbol{m}_{\mathbf{Cr}}^{\mathbf{UC.}}$ oriented in the out-of-plane direction at the interface owing to neighboring Co. Even though the conversion efficiency from orbital to spin might not be strong in Cr$_2$N, as in the case of pure Cr,[66] the spin would be scattered or transferred depending on the direction of $\boldsymbol{m}_{\mathbf{Cr}}^{\mathbf{UC.}}$, which is likely a spin-filtering effect by the $\boldsymbol{m}_{\mathbf{Cr}}^{\mathbf{UC.}}$ at interface.[27,47,69] Because of the antiferromagnetic coupling between the $M_{\text{Co/Pt}}$ and the $\boldsymbol{m}_{\mathbf{Cr}}^{\mathbf{UC.}}$, the polarized spin transferred through the interface is always opposite to the $M_{\text{Co/Pt}}$, resulting in field-free deterministic CIMS. To validate this, we examined the CIMS properties of the control sample with the 1-nm-thick Cu insertion between the Cr$_2$N layer and the [Co/Pt]$_3$ ferromagnetic multilayer, i.e., substrate//Cr$_2$N(5 nm)/Cu(1 nm)/[Co(0.35 nm)/Pt(0.3 nm)]$_3$/MgO(3 nm) [Fig. 6(b1)]. Due to weak SOI and non-magnetism of Cu, which can identify the effect of $\boldsymbol{m}_{\mathbf{Cr}}^{\mathbf{UC.}}$ on the observed field-free CIMS. Figures 6(b2) and 6(b3) show the representative out-of-plane AHE and the CIMS, respectively, where the charge current direction is parallel to the mirror symmetry ($m \parallel I$). Note that field-dependent CIMS was observed with the same polarity as the main sample [Fig. 2(c3)], while no field-free CIMS was evident by the Cu insertion. The result was comparable to the different configuration with the orthogonal charge current to the mirror symmetry axis as well (see Fig. S11 in the Supporting Information), which can be attributed to the absence of $\boldsymbol{m}_{\mathbf{Cr}}^{\mathbf{UC.}}$. We thus conclude that magnetic moment of Cr induced by the Co at the interface plays an essential role for field-free CIMS, which may become a key to elucidate one of the possible scenarios of the spin-filtering mechanism at the interface.

Some reports show that SOT-devices with Co/Pt multilayer exhibit CIMS by itself, without non-magnetic spin sources, which is referred to as self-induced SOT.[70] However, the self-induced SOT cannot become a major origin for the field-free CIMS in the present Cr$_2$N/[Co/Pt]$_3$ due to the following considerations. We examined the CIMS properties of the other control samples by replacing the Cr$_2$N layer with a Pt layer, as shown in Figs. 6(c1)-6(c3). Unlike the results for the main Cr$_2$N/[Co/Pt]$_3$, the polarity of CIMS was reversed: CW (CCW) for negative (positive) $H_x$, and no field-free CIMS was observed, although the [Co/Pt]$_3$ multilayer is consistent. This CIMS property is consistent with that observed in the conventional SOT-device such as Pt/CoPt bilayer systems, in which $y$-polarized spin current dominates the CIMS mechanisms.[57] These results suggest that the impact of spin source layer on SOT is much greater than that of Co/Pt multilayer itself. Furthermore, we confirmed the absence of field-free CIMS in the controlled [Co/Pt]$_6$ sample without Cr$_2$N layer as shown in Fig. 6(d1)-6(d3). We thus infer again that the impact of [Co/Pt]$_3$ for the field-free CIMS is minor, if any.

An interlayer exchange interaction owing to the $\boldsymbol{m}_{Cr}^{UC.}$ cannot contribute to the field-free CIMS, if any, in terms of the collinear magnetic structure of $\boldsymbol{m}_{Cr}^{UC.}$ and Co. It has been reported that the field-free CIMS can be observed in an SOT bilayer structure consisting of an antiferromagnetic layer with an in-plane Néel ventor and a perpendicularly magnetized ferromagnetic layer.[28] This is because the magnetic structure of ferromagnetic layer near the interface can be tilted to the in-plane direction through the interlayer exchange interaction, resulting in the effective in-plane magnetic field to break the inversion symmetry. Conversely, the Cr$_2$N has collinear antiferromagnetic structures with out-of-plane Néel ventors [Fig. 2(a)], which couples with perpendicularly magnetized Co/Pt ferromagnetic layer and the resultant magnetic structure cannot break the inversion symmetry. Therefore, field-free CIMS cannot be explained by the mechanims of interlayer exchange interaction for MXene-based SOT-devices.

It should be noted that non-centrosymmetric materials, such as monolayer TMDCs, exert out-of-plane SOTs in the adjacent ferromagnetic layer when the charge current flows orthogonal to the mirror axis, in which the sign of the out-of-plane SOT reverses with the sign of the charge current direction.[33,34,36] Recent theoretical study predicts a momentum-independent uniform spin configuration known as persistent spin texture for the other 2D non-centrosymmetric materials of CdTe and ZnTe,[71] which is expected to realize spintronic devices because of many advantages such as robustness against strain, layer thickness, and crystal distortion. In addition, the mechanism of the OHE that occurs in non-centrosymmetric materials has recently been examined as an intrinsic property.[72] All of these are classified into the symmetry-driven intrinsic mechanisms. In contrast to that, an unconventional out-of-plane SOT can emerge by the charge current flowing even in the direction parallel to the mirror axis, which is the specific characteristics of MXenes that cannot be explained by such the conventional scenario with inversion symmetry mentioned above. Therefore, investigation of the 2D-MXene/ferromagnet interface could be a key to facilitate the field-free CIMS in the SOT-device with the 2D-MXene.

The 2D-MXene has many advantages in principle, such as the bottom-up formability by the conventional sputtering, the phase stability up to ~650 °C, the sustainable light elements, and the process compatibility with CMOS technology. Furthermore, the charge-current-direction independent unconventional out-of-pane SOT may lead to a robust field-free CIMS for 2D SOT-MRAMs in the future.

## 3. Conclusion

We have reported the first demonstration of CIMS in the SOT-device with the sputter-deposited bare 2D-MXene of $Cr_2N$: substrate/$Cr_2N$/[Co/Pt]$_3$/MgO cap. The specific characteristics of the MXene-based SOT-device is the field-free CIMS via a possible out-of-plane unconventional SOT, regardless of the in-plane charge current direction with respect to crystal symmetry of $Cr_2N$, which is likely a robust field-free CIMS by MXene. A critical current density is $\sim 10^7$ A/cm$^2$, which is comparable to that of the conventional heavy-metal/ferromagnet systems. The $Cr_2N$ thickness dependence of SOT efficiency indicates the bulk OHE contribution, and $\sigma_{xz}^{\text{orb}(z)}$ dominates the OHE in $Cr_2N$ based on the first-principles calculations. The XMCD study indicates that the Cr of $Cr_2N$ is polarized by the adjacent ferromagnetic Co of Co/Pt multilayer, and the $m_{\text{Cr}}^{\text{UC.}}$ antiferromagnetically couples to the magnetic moment of Co. Thus, the field-free CIMS observed in MXene-based SOT-device can be predomonantly attributed to the bulk OHE and the spin-filtering-like effect at the interface due to $m_{\text{Cr}}^{\text{UC.}}$.

## 4. Experimental Section

### 4.1. Film fabrication and characterization

An $Al_2O_3$ crystal substrate with a (0001) plane orientation was cleaned with ethanol and acetone via ultrasonic cleaning and flash annealed at 650 °C for 30 min in the sputtering chamber with a base vacuum pressure of approximately $10^{-7}$ Pa. The $Cr_2N$ film was deposited on the substrate using the DC magnetron reactive sputtering for the Cr target at $T_{\text{sub}}$ from RT to 650 °C with the gas mixture of $N_2/(Ar + N_2) = 5\%$, where the deposition rate was 1.68 nm/min. The Co/Pt multilayer and MgO capping layer were deposited via DC and RF magnetron sputtering at RT. The crystal structure was investigated via X-ray diffraction (XRD; SmartLab; Rigaku Corporation) with Cu-$K_\alpha$ radiation. The surface roughness is evaluated via atomic force microscopy. The magnetic properties and anomalous Hall effect were measured at RT using a magnetic property measurement system (MPMS; Quantum Design Inc.) and a physical property measurement system (Dynacool; Quantum Design Inc.), respectively.

### 4.2. Element selective magnetic properties measured via XMCD

The XMCD measurements were performed at the BL14U Synchrotron Radiation Facility, NanoTerasu. Soft X-ray absorption spectra (XAS) were recorded using the total electron yield (TEY) method while scanning photon energy at RT. The XMCD signal was obtained by subtracting each XAS signal for circularly polarized light with positive and negative helicities. In particular, for Cr and N with tiny magnetic moments, the XAS measurement for each helicity was repeated five times and averaged to boost the signal-to-noise ratio. The magnetic field was applied perpendicularly to the surface of the sample. Element-selective magnetic properties against the applied field (ESMH) were measured for the $L_3$-edge of Co and Cr at RT.

### 4.3. CIMS and second-harmonic measurements

Photolithography and Ar ion milling were employed to fabricate the measurement devices with Hall cross and pillar patterns, in which the line width of charge current channel is 10 μm and the diameter of pillar is ~7 μm. A customized system was used for the CIMS experiments. A rectangular current pulse was applied to the current channel of the Hall cross devices with durations of 10 ms using a pulse generator (FG420; Yokogawa Electric Co.). The Hall voltage was recorded using a digital multimeter (7555; Yokogawa Electric Co.) at every interval between the current pulses, that is, 1 s after the last current pulse. The DC current to sense the Hall voltage was 0.5 mA (density: ~0.60 MA/cm$^2$) for Hall cross devices, and 0.2 mA (~0.14 MA/cm$^2$) for pillar devices, which were applied using a DC power source (G210, Yokogawa Electric Co.). The sensing current density was approximately 2 % of critical current density, which can be negligibly small for CIMS. The magnetic field from the electromagnet was uniform within a gap length of 3 cm and an area 5 cm in diameter. The device was placed away from the electromagnet for field-free CIMS to eliminate any residual field from the magnetic pole pieces. The second-harmonic Hall voltage was recorded using a lock-in amplifier (LI5640, NF Co.) while the in-plane applied field was scanned. A sinusoidal wave with an effective amplitude of 3 mA (density: ~4 MA/cm$^2$) and frequency of 33.123 Hz was applied using a pulse generator (FG420; Yokogawa Electric Co.). A common device and sample package were used for CIMS and second-harmonic measurements. All measurements were performed at RT.

### 4.4. Computational procedure for spin/orbital-Hall conductivities

First-principles calculations were performed using the Vienna ab initio Simulation Package.[73] Projector-augmented wave (PAW) pseudo-potentials were used for the atomic potentials of Cr and N with a plane-wave cut-off energy of 500 eV.[74] We adopted the generalized gradient approximation for the exchange and correlation energies, including the spin-orbit interaction with $10 \times 10 \times 10$ k-points in the first Brillouin zone.[75] We considered the onsite Coulomb interaction, $U = 3$ eV, for the Cr atom. The lattice parameters of Cr$_2$N are the same as those shown in Fig. 2(a).[76] The spin-Hall conductivity ($\sigma^{\text{spin}}$) and orbital-Hall conductivity ($\sigma^{\text{orb}}$) were calculated based on linear response theory as[77,78]:

$$\sigma_{\alpha\beta}^{X(\gamma)}(E) = \frac{e}{V} \sum_{\mathbf{k}} \Omega_{\alpha\beta}^{X(\gamma)}(\mathbf{k}, E). \tag{3}$$

$\Omega_{\alpha\beta}^{X(\gamma)}(\mathbf{k}, E)$ is the orbital Berry curvature provided by[79]:

$$\Omega_{\alpha\beta}^{X(\gamma)\gamma}(\mathbf{k},E) = 2\frac{\hbar^2}{m_e^2}\sum_{n>m}[f_{\mathbf{k}m}(E) - f_{\mathbf{k}n}(E)]\frac{\text{Im}\langle\mathbf{k}m|(\hat{p}_\alpha^{X(\gamma)})|\mathbf{k}n\rangle\langle\mathbf{k}n|\hat{p}_\beta|\mathbf{k}m\rangle}{(\varepsilon_{\mathbf{k}n} - \varepsilon_{\mathbf{k}m})^2}, \quad (4)$$

where $V$ denotes the unit-cell volume, $m_e$ denotes the electron mass, $m$ and $n$ denote the occupied and unoccupied band indices, respectively. $\hat{p}_\beta^{X(\gamma)}$ denotes "spin" or "orbital" current operator (X=spin or orbital), where $\hat{p}_\alpha^{\text{spin}(\gamma)} = \hat{p}_\alpha\hat{s}_\gamma + \hat{s}_\gamma\hat{p}_\alpha$ and $\hat{p}_\alpha^{\text{orbital}(\gamma)} = \hat{p}_\alpha\hat{L}_\gamma + L_\gamma\hat{p}_\alpha$. Furthermore, $\hat{p}_\alpha$ ($\hat{p}_\beta$) is $\alpha(\beta)$-axis component of the momentum operator, $\hat{s}_\gamma$ denotes the spin angular momentum operator with the spin quantum axis along $\gamma$ direction, and $\hat{L}_\gamma$ denotes the orbital angular momentum operator along $\gamma$ direction. Additionally, $|\mathbf{k}n\rangle$ denotes the eigenstate with the eigenenergy $\varepsilon_{\mathbf{k}n}$, and $f_{\mathbf{k}n}(E)$ denotes the occupation function for band $n$ and wave-vector $\mathbf{k}$ at the energy ($E$) relative to the Fermi level ($E_F$). The $\sigma^{\text{spin(orb)}}$ of $Cr_2N$ was computed using $30 \times 30 \times 30$ k points in the first Brillouin zone.

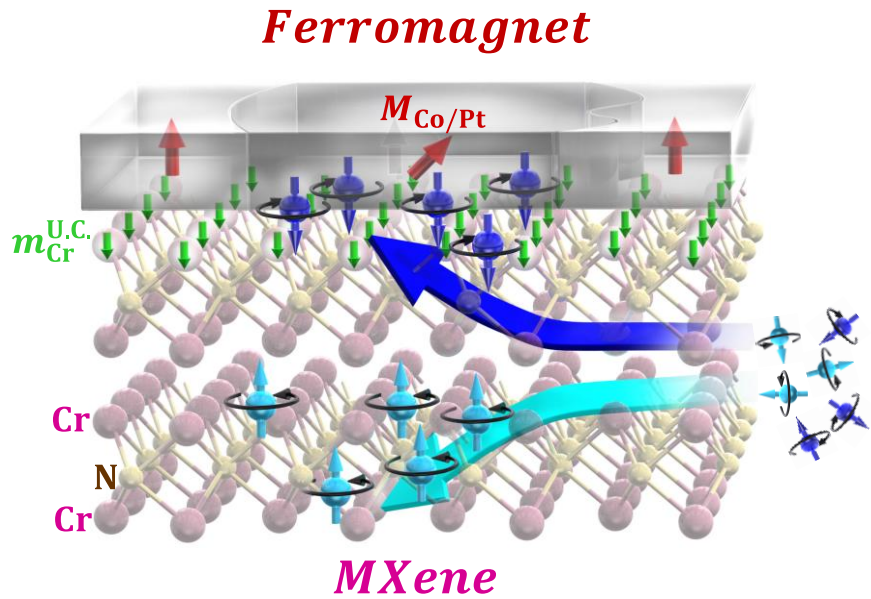

**Figure 1.** Concept of the field-free CIMS in the two-dimensional MXene-based SOT-device, i.e., the MXene ($Cr_2N$)/ferromagnet bilayer system. The electron-spins oriented in the out-of-plane direction emerge through the pronounced orbital Hall effect in the MXene layer. The thick green arrows at the interface represent the out-of-plane uncompensated magnetic moment of Cr ($m_{Cr}^{U.C.}$) induced by the adjacent ferromagnetic layer. The $m_{Cr}^{U.C.}$ can act as a spin-filter that transfers the polarized electron-spins with the same orientation as the $m_{Cr}^{U.C.}$. The out-of-plane oriented spins exert the torque on the ferromagnetic layer, resulting in the magnetic domain switching without in-plane magnetic fields.

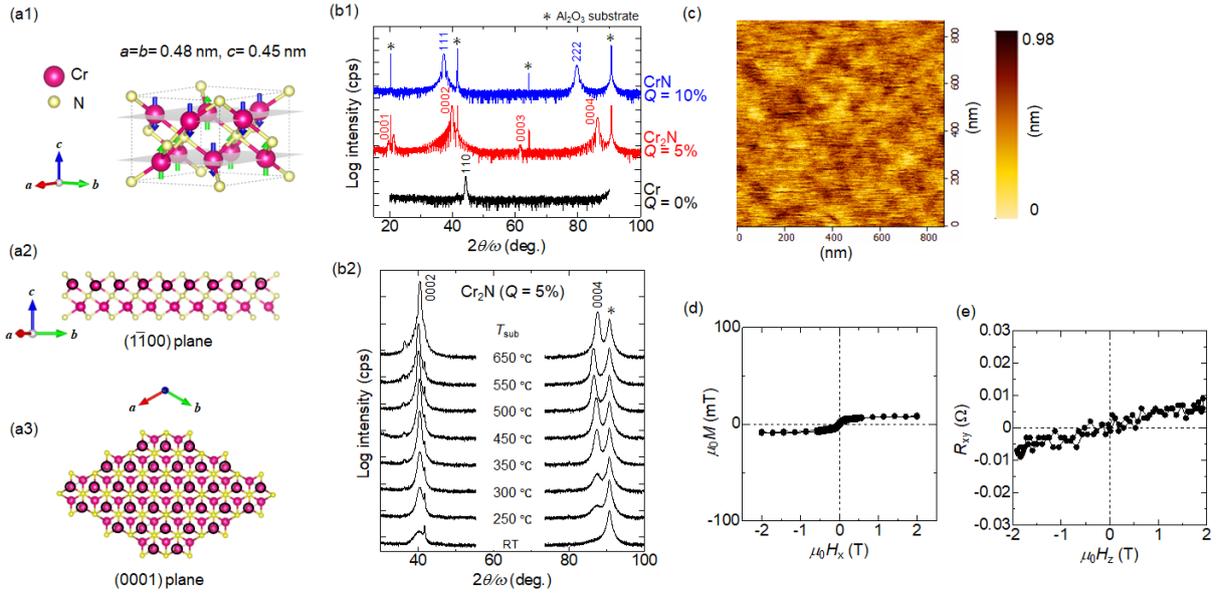

**Figure 2.** (a1) Unit cell model of the Cr$_2$N bare MXene together with the possible magnetic structure predicted by first-principles calculation. (a2, a3) Cross-sectional view (a2) and plane view (a3) for the 3 × 3 × 1 supercell, where the Cr atoms at the top layer are surrounded by black circles. (b1, b2) Out-of-plane XRD profiles for the 20-nm-thick Cr-N films deposited with different N$_2$ flow ratio $Q = $ N$_2$/(Ar+N$_2$) while reactive sputtering deposition (b1), and substrate temperature ($T_{sub}$) (b2). (c) Atomic force microscopy image for the 20-nm-thick Cr$_2$N film. (d) Magnetization curve measured by the in-plane magnetic field ($H_x$). (e) Anomalous Hall resistance as a function of out-of-plane magnetic field ($H_z$).

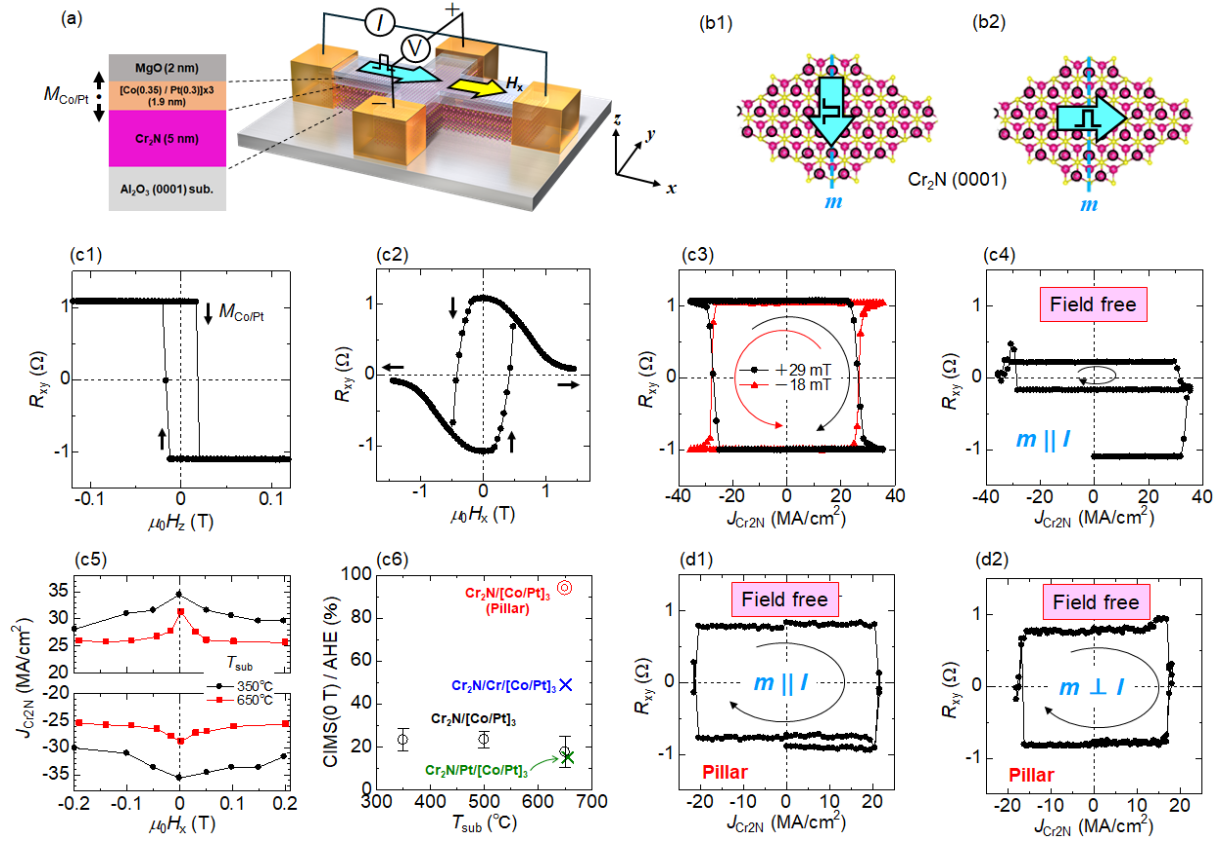

**Figure 3.** (a) Measurement configuration of the current-induced magnetization switching (CIMS) and representative stacking structure, where the 5-nm-thick MXene layer consists of ~10 unit-layer of $Cr_2N$. (b1, b2) Two different directions of the current pulse with respect to mirror symmetry line (*m*) in the CIMS demonstration. (c1, c2) Anomalous Hall resistance for the same sample with magnetic field along out-of-plane ($H_z$) and in-plane directions. (c3, c4) Representative CIMS with and without $H_x$. (c5) $H_x$ dependence of $J_{Cr_2N}$ for the $Cr_2N$ growth temperature ($T_{sub}$) of 350 °C and 650 °C. (c6) Partial field-free CIMS ratio relative to the full switching by $H_z$ for various $T_{sub}$. The green and blue symbols represent the same results, but with the insertion of 1-nm-thick Cr and Pt layers, respectively, for comparison. The red symbol indicates the result with circular shaped pillar devices. (d1, d2) Representative field-free CIMS loops with and without $H_x$ for the pillar device (see Fig. S5 in the Supporting Information).

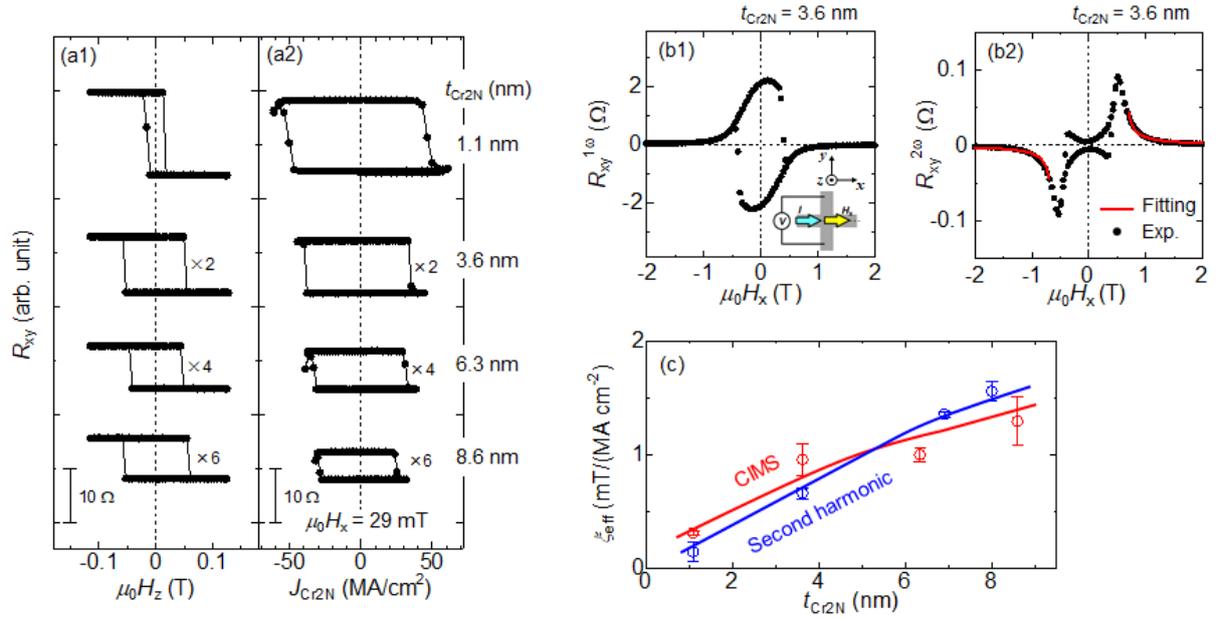

**Figure 4.** (a1, a2) Cr$_2$N layer thickness ($t_{Cr2N}$) dependence of the out-of-plane AHE and CIMS loops with the bias field $\mu_0 H_x = +29$ mT, for the SOT-device of substrate//Cr$_2$N($t_{Cr2N}$)/[Co(0.35 nm)/Pt(0.3 nm)]$_3$/MgO(2 nm). (b1, b2) High-field in-plane AHE loops for the first- (1ω) and second-harmonic (2ω) Hall resistance. The red curve in Fig. 4(b2) represents the fitting result using Eq. (2) to distinguish the damping-like ($H_{DL}$) and field-like ($H_{FL}$) SOT effective fields from the magneto thermoelectric effect. (c) Damping-like SOT efficiency ($\xi_{eff}$) as a function of $t_{Cr2N}$ estimated by both the critical $J_{Cr2N}$ in CIMS loops (red) and the $H_{DL}$ value in the second-harmonic measurements (blue).

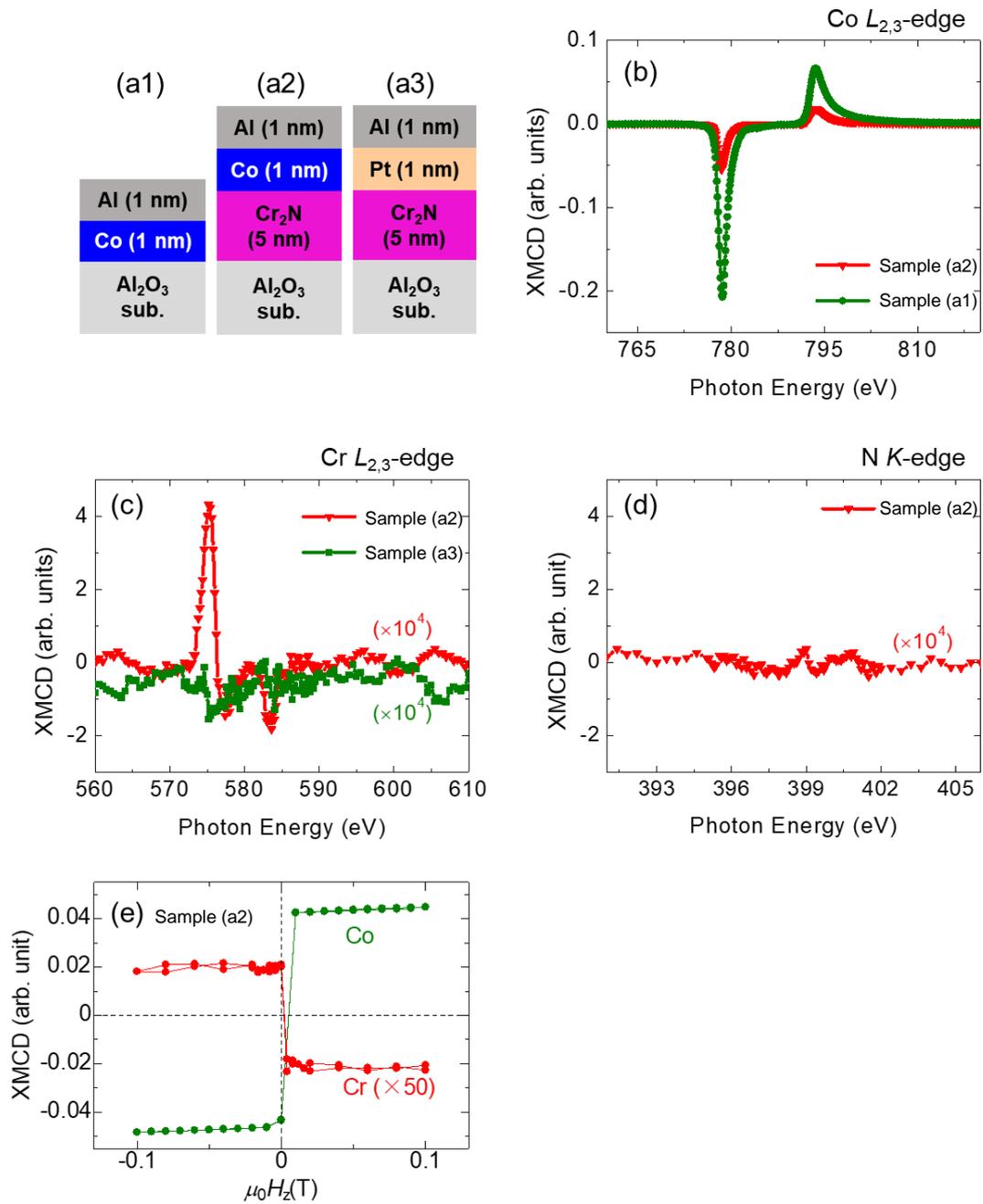

**Figure 5.** (a1-a3) Stacking structures for the XMCD measurement at the synchrotron radiation facility. (b-d) XMCD spectra for the Co $L_{2,3}$-edge (b), Cr $L_{2,3}$-edge (c), and N $K$-edge (d) of the samples shown in Fig. 5(a1-a3). (e) Element-selective out-of-plane magnetic properties for Co and Cr of the sample (a2).

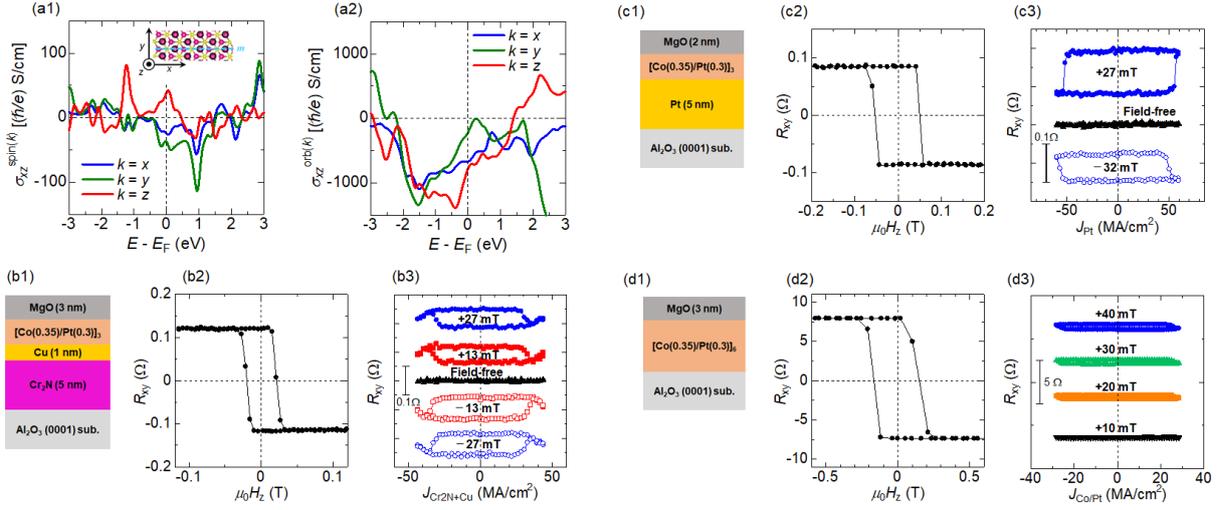

**Figure 6.** (a) Energy dependent spin-Hall conductivity ($\sigma_{xz}^{\text{spin}(k)}$) and orbital-Hall conductivity ($\sigma_{xz}^{\text{orb}(k)}$) for the Cr$_2$N, where $x$, $z$, and $k$ represent the directions of charge current, spin current, and spin/orbital polarization, respectively. (b-d) CIMS results for the control Hall-cross samples, with 1-nm-thick Cu layer insertion and parallel charge current to mirror symmetry axis (b), with Pt underlayer (c), and with only [Co/Pt]$_6$ multilayer (d).

**Supporting Information**

Supporting Information is available from the Wiley Online Library or from the author.

**Acknowledgements**

The authors would like to thank Dr. Shiratsuchi for the discussion about XMCD results. The XMCD measurements were performed at the BL14U of the synchrotron radiation facility NanoTerasu. This work was supported by KAKENHI Grants-in-Aid No. 23K22803 from the Japan Society for the Promotion of Science (JSPS). Part of this work was performed under the Cooperative Research Project Program of the RIEC, Tohoku University.

Received: ((will be filled in by the editorial staff))
Revised: ((will be filled in by the editorial staff))
Published online: ((will be filled in by the editorial staff))

**References**

[1] A. K. Geim, S. V. Morozov, D. Jiang, Y. Zhang, S. V. Dubonos, I. V. Grigorieva, A. A. Firsov and K. S. Novoselov. Electric Field Effect in Atomically Thin Carbon Films. *Science* **2004**, 306, 666.

[2] Q. H. Wang, K. K.-Zadeh, A. Kis, J. N. Coleman, and M. S. Strano. Electronics and optoelectronics of two-dimensional transition metal dichalcogenides. Nature Nanotech. **2012**, 7, 699.

[3] B. Radisavljevic, A. Radenovic, J. Brivio, V. Giacometti, and A. Kis. Single-layer $MoS_2$ transistors. *Nature Nanotech*. **2011**, 6, 147.

[4] H. Fang, S. Chuang, T. C. Chang, K. Takei, T. Takahashi, and A. Javey. High-Performance Single Layered $WSe_2$ p-FETs with Chemically Doped Contacts. *Nano Lett*. **2012**, 12, 3788.

[5] B. Radisavljevic, M. B. Whitwick, and A. Kis. Integrated Circuits and Logic Operations Based on Single-Layer $MoS_2$. *ACS Nano*. **2011**, 12, 9934.

[6] A. Pospischil, M. M. Furchi, and T. Mueller. Solar-energy conversion and light emission in an atomic monolayer p–n diode. *Nature Nanotech*. **2014**, 9, 257.

[7] A. Splendiani, L. Sun, Y. Zhang, T. Li, J. Kim, C.-Y. Chim, G. Galli, and F. Wang. Emerging Photoluminescence in Monolayer$MoS_2$. *Nano Lett*. **2010**, 10, 1271.

[8] Y. Cao, V. Fatemi, S. Fang, K. Watanabe, T. Taniguchi, E. Kaxiras, and P. J.-Herrero. Unconventional superconductivity in magic-angle graphene superlattices. *Nature* **2017**, 556, 43.

[9] Y. Qi, M. A. Sadi, D. Hu, M. Zheng, Z. Wu, Y. Jiang, and Y. P. Chen. Recent Progress in Strain Engineering on Van der Waals 2D Materials: Tunable Electrical, Electrochemical, Magnetic, and Optical Properties. *Adv. Mater*. **2023**, 35, 2205714.

[10] M. Naguib, M. Kurtoglu, V. Presser, J. Lu, J. Niu, M. Heon, L. Hultman, Y. Gogotsi, and M. W. Barsoum. Two-Dimensional Nanocrystals Produced by Exfoliationof $Ti_3AlC_2$. *Adv. Mater*. **2011**, 23, 4248.

[11] M Naguib, V. N. Mochalin, M. W. Barsoum, and Y. Gogotsi. Two-Dimensional Materials: 25th Anniversary Article: MXenes: A New Family of Two-Dimensional Materials. *Adv. Mater*. **2014**, 26, 992.

[12] X. Li, M. Li, Z. Huang, G. Liang, Z. Chen, Q. Yang, Q. Huang, and C. Zhi. Activating the $I^0/I^+$ redox couple in an aqueous $I_2$–Zn battery to achieve a high voltage plateau. *Energy Environ. Sci*. **2021**, 14, 407.

[13] S. Isogami, and Y. K. Takahashi. Antiperovskite Magnetic Materials with 2p Light Elements for Future Practical Applications. *Adv. Electron. Mater*. **2023**, 9, 2200515.

[14] N. Driscoll, A. G. Richardson, K. Maleski, B. Anasori, O. Adewole, P. Lelyukh, L. Escobedo, D. K. Cullen, T. H. Lucas, Y. Gogotsi, and F. Vitale. Two-Dimensional $Ti_3C_2$ MXene for High-Resolution Neural Interfaces. *ACS Nano* **2018**, 12, 10419.

[15] J. Zhang, N. Kong, S. Uzun, A. Levitt, S. Seyedin, P. A. Lynch, S. Qin, M. Han, W. Yang, J. Liu, X. Wang, Y. Gogotsi, and J. M. Razal. Scalable Manufacturing of Free-Standing, Strong $Ti_3C_2T_x$ MXene Films with Outstanding Conductivity. *Adv. Mater.* **2020**, 32, 2001093.

[16] S. Ahn, T.-H. Han, K. Maleski, J. Song, Y.-H. Kim, M.-H. Park, H. Zhou, S. Yoo, Y. Gogotsi, and T.-W. Lee. A 2D Titanium Carbide MXene Flexible Electrode for High-Efficiency Light-Emitting Diodes. *Adv. Mater.* **2020**, 32, 2000919.

[17] S. Zhao, X. Meng, K. Zhu, F. Du, G. Chen, Y. Wei, Y. Gogotsi, and Y. Gao. Li-ion uptake and increase in interlayer spacing of $Nb_4C_3$ MXene. *Energy Storage Materials* **2017**, 8, 42.

[18] H. Yang, S. O. Valenzuela, M. Chshiev, S. Couet, B. Dieny, B. Dlubak, A. Fert, K. Garello, M. Jamet, D.-E. Jeong, K. Lee, T. Lee, M.-B. Martin, G. S. Kar, P. Sénéor, H.-J. Shin, and S. Roche. Two-dimensional materials prospects for non-volatile spintronic memories. *Nature* **2022**, 606, 663.

[19] Y. Liu and Q. Shao. Two-Dimensional Materials for Energy Efficient Spin−Orbit Torque Devices. ACS Nano **2020**, 14, 9389.

[20] S. Maekawa, T. Kikkawa, H. Chudo, J. Ieda, E. Saitoh. Spin and spin current—From fundamentals to recent progress. *J. Appl. Phys.* **2023**, 133, 020902.

[21] V. D. Nguyen, S. Rao, K. Wostyn, and S. Couet. Recent progress in spin-orbit torque magnetic random-access memory. *npj Spintronics* **2024**, 2, 48.

[22] S. Fukami, T. Anekawa, C. Zhang, and H. Ohno. A spin–orbit torque switching scheme with collinear magnetic easy axis and current configuration. *Nature Nanotech.* **2016**, 11, 621.

[23] H. Wu, J. Zhang, B. Cui, S. A. Razavi, X. Che, Q. Pan, D. Wu, G. Yu, X. Han, and K. L. Wang. Field-free approaches for deterministic spin–orbit torque switching of the perpendicular magnet. *Mater. Futures* **2022**, 1, 022201.

[24] T.-Y. Chen, H.-I Chan, W.-B. Liao, and C.-F. Pai. Current-Induced Spin-Orbit Torque and Field-Free Switching in Mo-Based Magnetic Heterostructures. *Phsy. Rev. Appl.* **2018**, 10, 044038.

[25] Z. Zheng, Y. Zhang, V. L.-Dominguez, L. S.-Tejerina, J. Shi, X. Feng, L. Chen, Z. Wang, Z. Zhang, K. Zhang, B. Hong, Y. Xu, Y. Zhang, M. Carpentieri, A. Fert, G. Finocchio, W. Zhao, and P. K. Amiri. Field-free spin-orbit torque-induced switching of perpendicular

magnetization in a ferrimagnetic layer with a vertical composition gradient. *Nature Communications* **2021**, 12, 4555.

[26] V. P. Amin, J. Zemen, and M. D. Stiles. Interface-Generated Spin Currents. *Phys. Rev. Lett*. **121**, 136805 (2018).

[27] G. Choi, J. Ryu, S. Lee, J. Kang, N. Noh, J. M. Yuk, B.-G. Park. Thickness Dependence of Interface-Generated Spin Currents in Ferromagnet/Ti/CoFeB Trilayers. *Adv. Mater. Int.* **2022**, 9, 2201317.

[28] S. Fukami, C. Zhang, S. DuttaGupta, A. Kurenkov, and H. Ohno. Magnetization switching by spin–orbit torque in an antiferromagnet–ferromagnet bilayer system. *Nature Mater.* **2016**, 15, 535.

[29] Y.-C. Lau, D. Betto, K. Rode, J. M. D. Coey, and P. Stamenov. Spin–orbit torque switching without an external field using interlayer exchange coupling. *Nature Nanotech.* **2016**, 11, 758.

[30] Y. You, H. Bai, X. Feng, X. Fan, L. Han, X. Zhou, Y. Zhou, R. Zhang, T. Chen, F. Pan, and C. Song. Cluster magnetic octupole induced out-of-plane spin polarization in antiperovskite antiferromagnet. *Nature Communications* **2021**, 12, 6524.

[31] S. Hu, D.-F. Shao, H. Yang, C. Pan, Z. Fu, M. Tang, Y. Yang, W. Fan, S. Zhou, E. Y. Tsymbal, and X. Qiu. Efficient perpendicular magnetization switching by a magnetic spin Hall effect in a noncollinear antiferromagnet. *Nature Communications* **2022**, 13, 4447.

[32] C. Cao, S. Chen, R.-C. Xiao, Z. Zhu, G. Yu, Y. Wang, X. Qiu, L. Liu, T. Zhao, D.-F. Shao, Y. Xu, J. Chen, and Q. Zhan. Anomalous spin current anisotropy in a noncollinear antiferromagnet. *Nature Communications* **2023**, 14, 5873.

[33] D. Macneill, G. M. Stiehl, M. H. D. Guimaraes, R. A. Buhrman, J. Park, D. C. Ralph. Control of spin–orbit torques through crystal symmetry in WTe$_2$/ferromagnet bilayers. *Nat. Phys*. **2017**, 13, 300.

[34] I.-H. Kao, R. Muzzio, H. Zhang, M. Zhu, J. Gobbo, S. Yuan, D. Weber, R. Rao, J. Li, J. H. Edgar, J. E. Goldberger, J. Yan, D. G. Mandrus, J. Hwang, R. Cheng, J. Katoch, and S. Singh. Deterministic switching of a perpendicularly polarized magnet using unconventional spin–orbit torques in WTe$_2$. *Nature Materials* **2022**, 21, 1029.

[35] S. N. Kajale, T. Nguyen, N. T. Hung, M. Li, D. Sarkar. Field-free deterministic switching of all–van der Waals spin- orbit torque system above room temperature. *Sci. Adv*. **2024**, 10, 8669.

[36] X. Wang, H. Wu, R. Qiu, X. huang, J. Zhang, J. long, Y. Yao, Y. Zhao, Z. Zhu, J. Wang, S. Shi, H. Chang, W. Zhao. Room temperature field-free switching of CoFeB/MgO heterostructure based on large-scale few-layer WTe$_2$. *Cell Rep. Phys. Sci*. **2023**, 4, 101468.

[37] I. Shin, W. J. Cho, E.-S. An, S. Park, H.-W. Jeong, S. Jang, W. J. Baek, S. Y. Park, D.-H. Yang, J. H. Seo, G.-Y. Kim, M. N. Ali, S.-Y. Choi, H.-W. Lee, J. S. Kim, S. D. Kim, and G.-H. Lee. Spin–Orbit Torque Switching in an All-Van der Waals Heterostructure. *Adv. Mater*. **2022**, 34, 2101730.

[38] E. Grimaldi, V. Krizakova, G. Sala, F. Yasin, S. Couet, G. S. Kar, K. Garello, and P. Gambardella. Single-shot dynamics of spin–orbit torque and spin transfer torque switching in three-terminal magnetic tunnel junctions. *Nature Nanotech.* **2020**, 15, 111.

[39] N. Sato, 2020 Symposia on VLSI Technology and Circuits, Honolulu (Virtual), June **2020**.

[40] S. Khan, A. Mahmood, A. Shah, Q. Raza, M. A. Rasheed, and I. Ahmad. Structural and optical analysis of Cr$_2$N thin films prepared by DC magnetron sputtering. *Int. J. Miner. Metall. Mater*. **2015**, 22, 197.

[41] M. A. Gharavi, G. Greczynski, F. Eriksson, J. Lu, B. Balke, D. Fournier, A. le Febvrier, C. Pallier, and P. Eklund. Synthesis and characterization of single-phase epitaxial Cr$_2$N thin films by reactive magnetron sputtering. *J. Mater. Sci.* **2019**, 54, 1434.

[42] J. Kim, D. Go, H. Tsai, D. Jo, K. Kondou, H.-W. Lee, and Y. Otani. Nontrivial torque generation by orbital angular momentum injection in ferromagnetic-metal/Cu/Al$_2$O$_3$ trilayers. *Phys. Rev. B* **2021**, 103, L020407.

[43] S. Ding, A. Ross, D. Go, L. Baldrati, Z. Ren, F. Freimuth, S. Becker, F. Kammerbauer, J. Yang, G. Jakob, Y. Mokrousov, and M. Kläui. Harnessing orbital-to-spin conversion of interfacial orbital currents for efficient spin-orbit torques. *Phys. Rev. Lett.* **2020**, 125, 177201.

[44] T. An, B. Cui, M. Zhang, F. Liu, S. Cheng, K. Zhang, X. Ren, L. Liu, B. Cheng, C. Jiang, and J. Hu. Electrical Manipulation of Orbital Current via Oxygen Migration in Ni$_{81}$Fe$_{19}$/CuO$_x$/TaN Heterostructure. *Adv. Mater*. **2023**, 35, 2300858.

[45] Y. Zhang, X. Ren, R. Liu, Z. Chen, X. Wu, J. Pang, W. Wang, G. Lan, K. Watanabe, T. Taniguchi, Y. Shi, G. Yu, and Q. Shao. Robust Field-Free Switching Using Large Unconventional Spin-Orbit Torque in an All-Van der Waals Heterostructure. *Adv. Mater.* **2024**, 36, 2406464.

[46] Y. Dai, J. Xiong, Y. Ge, B. Cheng, L.Wang, P. Wang, Z. Liu, S. Yan, C. Zhang, X. Xu, Y. Shi, S.-W. Cheong, C. Xiao, S. A. Yang, S.-J. Liang, and F. Miao. Interfacial magnetic

spin Hall effect in van der Waals Fe3GeTe2/MoTe2 heterostructure. *Nature Communications* **2024**, 15, 1129.

[47] T. Song, X. Cai, M. W.-Y. Tu, X. Zhang, B. Huang, N. P. Wilson, K. L. Seyler, L. Zhu, T. Taniguchi, K. Watanabe, M. A. McGuire, D. H. Cobden, D. Xiao, W. Yao, X. Xu. Giant tunneling magnetoresistance in spin-filter van der Waals heterostructures. *Science* **2018**, 360, 1214.

[48] J. D. Browne, P. R. Liddell, R. Street, and T. Mills. An Investigation of the Antiferromagnetic Transition of CrN. *Phys. Stat. Sol. (a)* **1970**, 1, 715.

[49] G. Wang. Theoretical Prediction of the Intrinsic Half-Metallicity in Surface Oxygen-Passivated $Cr_2N$ MXene. *J. Phys. Chem. C* **2016**, 120, 18850.

[50] H. Kumar, N. C. Frey, L. Dong, and V. B. Shenoy. Tunable Magnetism and Transport Properties in Nitride MXenes. *ACS Nano* **2017**, 11, 7648.

[51] S. J. G.-Ojeda, R. P.-Pérez, D. M.-Lopez, D. M. Hoat, J. G.-Sánchez, M. G. M.-Armenta. Strain Effects on the Two-Dimensional $Cr_2N$ MXene: An Ab Initio Study. *ACS Omega* **2022**, 7, 33884.

[52] S. J. G.-Ojeda, R. P.-Pérez, J. G.-Sánchez, M. G. M.-Armenta. MXene heterostructures based on $Cr_2C$ and $Cr_2N$: evidence of strong interfacial interactions that induce an antiferromagnetic alignment. *Graphene and 2D Materials* **2024**, 9, 47.

[53] S. Isogami, Y. Shiokawa, A. Tsumita, E. Komura, Y. Ishitani, K. Hamanaka, T. Taniguchi, S. Mitani, T. Sasaki, and M. Hayashi. Spin–orbit torque driven magnetization switching in W/CoFeB/MgO-based type-Y three terminal magnetic tunnel junctions. *Sci. Rep.* **2022**, 11, 16676.

[54] G. J. Lim, W. L. Gan, W. C. Law, C. Murapaka, W. S. Lew. Spin-orbit torque induced multi-state magnetization switching in Co/Pt hall cross structures at elevated temperatures. *J. Magn. Magn. Mater.* **2020**, 514, 167201.

[55] O. J. Lee, L. Q. Liu, C. F. Pai, Y. Li, H. W. Tseng, P. G. Gowtham, J. P. Park, D. C. Ralph, and R. A. Buhrman. Central role of domain wall depinning for perpendicular magnetization switching driven by spin torque from the spin Hall effect. *Phys. Rev. B* **2014**, 89, 024418.

[56] X. Qiu, W. Legrand, P. He, Y. Wu, J. Yu, R. Ramaswamy, A. Manchon, and H. Yang. Enhanced Spin-Orbit Torque via Modulation of Spin Current Absorption. *Phys. Rev. Lett.* **2016**, 117, 217206.

[57] C. O. Avci, K. Garello, M. Gabureac, A. Ghosh, A. Fuhrer, S. F. Alvarado, and P. Gambardella. Interplay of spin-orbit torque and thermoelectric effects in ferromagnet/normal-metal bilayers. *Phys. Rev. B* **2014**, 90, 224427.

[58] B. T. Thole, P. Carra, F. Sette, and G. van der Laan. X-ray circular dichroism as a probe of orbital magnetization. *Phys. Rev. Lett.* **1992**, 68, 1943.

[59] P. Carra, B.T. Thole, M. Altarelli, and X. Wang. X-Ray Circular Dichroism and Local Magnetic Fields. *Phys. Rev. B* **1993**, 70, 694.

[60] O. Hjortstam, J. Trygg, J. M. Wills, B. Johansson, and O. Eriksson. Calculated spin and orbital moments in the surfaces of the 3d metals Fe, Co, and Ni and their overlayers on Cu(001). *Phys. Rev. B* **1996**, 53, 9204.

[61] Y. Shiratsuchi, H. Noutomi, H. Oikawa, T. Nakamura, M. Suzuki, T. Fujita, K. Arakawa, Y. Takechi, H. Mori, T. Kinoshita, M. Yamamoto, and R. Nakatani. Detection and In Situ Switching of Unreversed Interfacial Antiferromagnetic Spins in a Perpendicular-Exchange-Biased System. *Phys. Rev. Lett*. **2012**, 109, 077202.

[62] C. S.-Hanke, R. G.-Arrabal, J. E. Prieto, E. Andrzejewska, N. Gordillo, D. O. Boerma, R. Loloee, J. Skuza, R. A. Lukaszew. Observation of nitrogen polarization in Fe-N using soft x-ray magnetic circular dichroism. *J. Appl. Phys.* **2006**, 99, 08B709.

[63] J. Bass, W. P Pratt Jr. Spin-diffusion lengths in metals and alloys, and spin-flipping at metal/metal interfaces: an experimentalist's critical review. *J. Phys.: Condens. Matter.* **2007**, 19, 183201.

[64] D. Qu, S. Y. Huang, and C. L. Chien. Inverse spin Hall effect in Cr: Independence of antiferromagnetic ordering. *Phys. Rev. B* **2015**, 92, 020418(R).

[65] B. A. Bernevig, T. L. Hughes, and S.-C. Zhang. Orbitronics : The Intrinsic Orbital Current in p-Doped Silicon. *Phys. Rev. Lett.* **2005**, 95, 066601.

[66] S. Lee, M.-G. Kang, D. Go, D. Kim, J.-H. Kang, T. Lee, G.-H. Lee, J. Kang, N. J. Lee, Y. Mokrousov, S. Kim, K.-J. Lee, B.-G. Park. Efficient conversion of orbital Hall current to spin current for spin-orbit torque switching. *Comm. Phys*. **2021**, 4, 234.

[67] G. Sala, and P. Gambardella. Giant orbital Hall effect and orbital-to-spin conversion in 3d, 5d, and 4f metallic heterostructures. *Phys. Rev. Res.* **2022**, 4, 033037.

[68] D. Jo, D. Go, and H.-W. Lee. Gigantic intrinsic orbital Hall effects in weakly spin-orbit coupled metals. *Phys. Rev. B* **2018**, 98, 214405.

[69] V. P. Amin, P. M. Haney, and M. D. Stiles. Interfacial spin–orbit torques. *J. Appl. Phys.* **2020**, 128, 151101.

[70] B. Jinnai, C. Zhang, A. Kurenkov, M. Bersweiler, H. Sato, S. Fukami, and H. Ohno. Spin-orbit torque induced magnetization switching in Co/Pt multilayers. *Appl. Phys. Lett.* **2017**, 111, 102402.

[71] M. K. Mohanta, and P. Jena. Symmetry-driven persistent spin texture for the two-dimensional nonsymmorphic CdTe and ZnTe crystal structures. *Phys Rev. B* **2023**, 108, 085432.

[72] S. Bhowal, and S. Satpathy. Intrinsic orbital moment and prediction of a large orbital Hall effect in two-dimensional transition metal dichalcogenides. *Phys. Rev. B* **2020**, 101, 121112(R).

[73] G. Kresse and J. Furthmüller. Efficient iterative schemes for ab initio total-energy calculations using a plane-wave basis set. *Phys. Rev. B* **1996**, 54, 11169.

[74] P. E. Blöchl. Projector augmented-wave method. *Phys. Rev. B* **1994**, 50, 17953.

[75] J. P. Perdew, K. Burke, and M. Ernzerhof. Generalized gradient approximation made simple. *Phys. Rev. Lett.* **1996**, 77, 3865.

[76] V. I. Anisimov, J. Zaanen, and O. K. Andersen. Band theory and Mott insulators: Hubbard U instead of Stoner. *Phys. Rev. B* **1991**, 44, 943.

[77] H. Nakano. A method of calculation of electrical conductivity. *Prog. Theor. Phys.* **1956**, 15, 77.

[78] R. Kubo. Statistical mechanical theory of irreversible processes: General theory and simple applications in magnetic and con duction problems. *J. Phys. Soc. Jpn.* **1957**, 12, 570.

[79] Y. Miura, and K. Masuda. First-principles calculations on the spin anomalous Hall effect of ferromagnetic alloys. *Phys. Rev. Mater.* **2021**, 5, L101402.